\documentclass[prb,preprint,amsmath,amssymb]{revtex4}
\begin{document}
\author{Kevin Leung$^*$}
\affiliation{Sandia National Laboratories, MS 1415,
Albuquerque, NM 87185\\
$^*${\tt kleung@sandia.gov}}
\date{\today}
\title{Electronic Structure Modeling of Electrochemical Reactions at
Electrode/Electrolyte Interfaces in Lithium Ion Batteries}

\input epsf
 
\begin{abstract}
 
We review recent {\it ab initio} molecular dynamics studies of
electrode/electrolyte interfaces in lithium ion batteries.  Our goals
are to introduce experimentalists to simulation techniques applicable to models
which are arguably most faithful to experimental conditions so far, and to
emphasize to theorists that the inherently interdisciplinary nature of this
subject requires bridging the gap between solid and liquid state perspectives.
We consider liquid ethylene carbonate (EC) decomposition on lithium intercalated
graphite, lithium metal, oxide-coated graphite, and spinel manganese oxide
surfaces.  These calculations are put in the context of more widely studied
water-solid interfaces.  Our main themes include kinetically controlled
two-electron-induced reactions, the breaking of a previously much neglected
chemical bond in EC, and electron tunneling.  Future work on modeling
batteries at atomic lengthscales requires capabilities beyond 
state-of-the-art, which emphasizes that applied battery research can and
should drive fundamental science development.
\vspace*{0.1in}\\
keywords: solid electrolyte interphase; {\it ab initio molecular dynamics};
lithium manganese oxide; ethylene carbonate; electron transfer
 
\end{abstract}
 
\maketitle
 
\section{Introduction: background and context}
\label{introduction}

\begin{figure}
\centerline{\hbox{ (a) \epsfxsize=3.00in \epsfbox{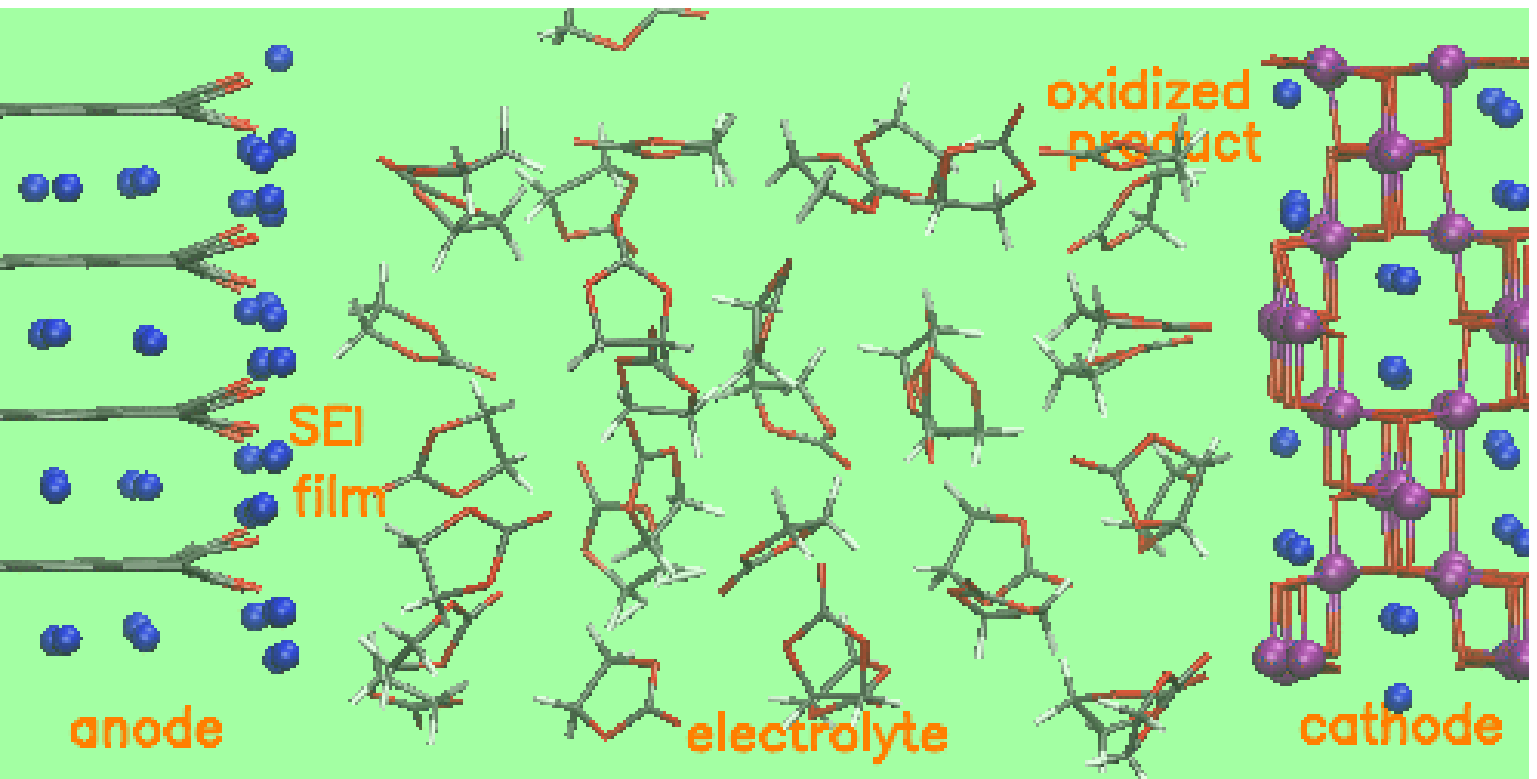} }}
\centerline{\hbox{ (b) \epsfxsize=1.425in \epsfbox{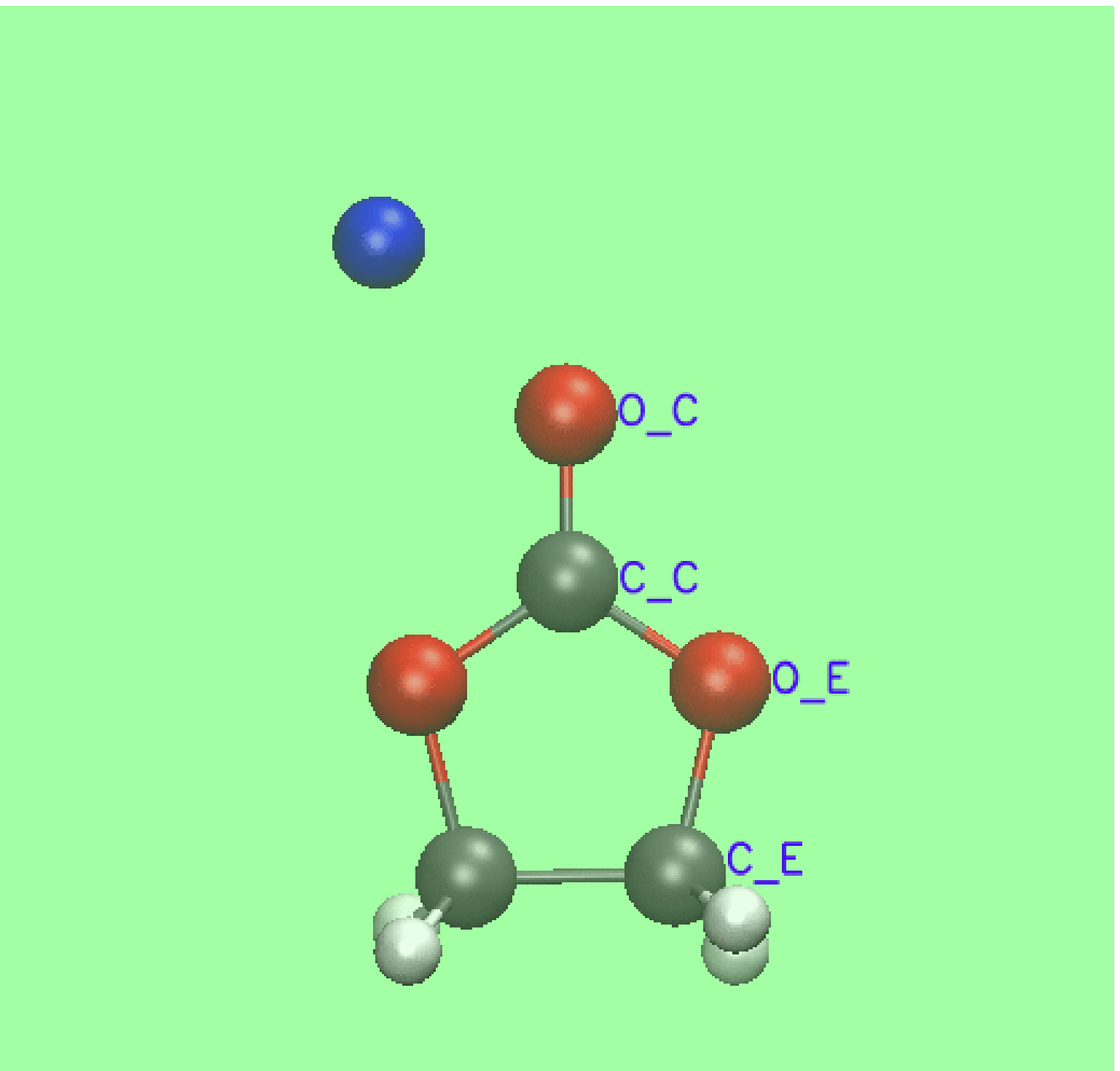} 
                   (c) \epsfxsize=1.425in \epsfbox{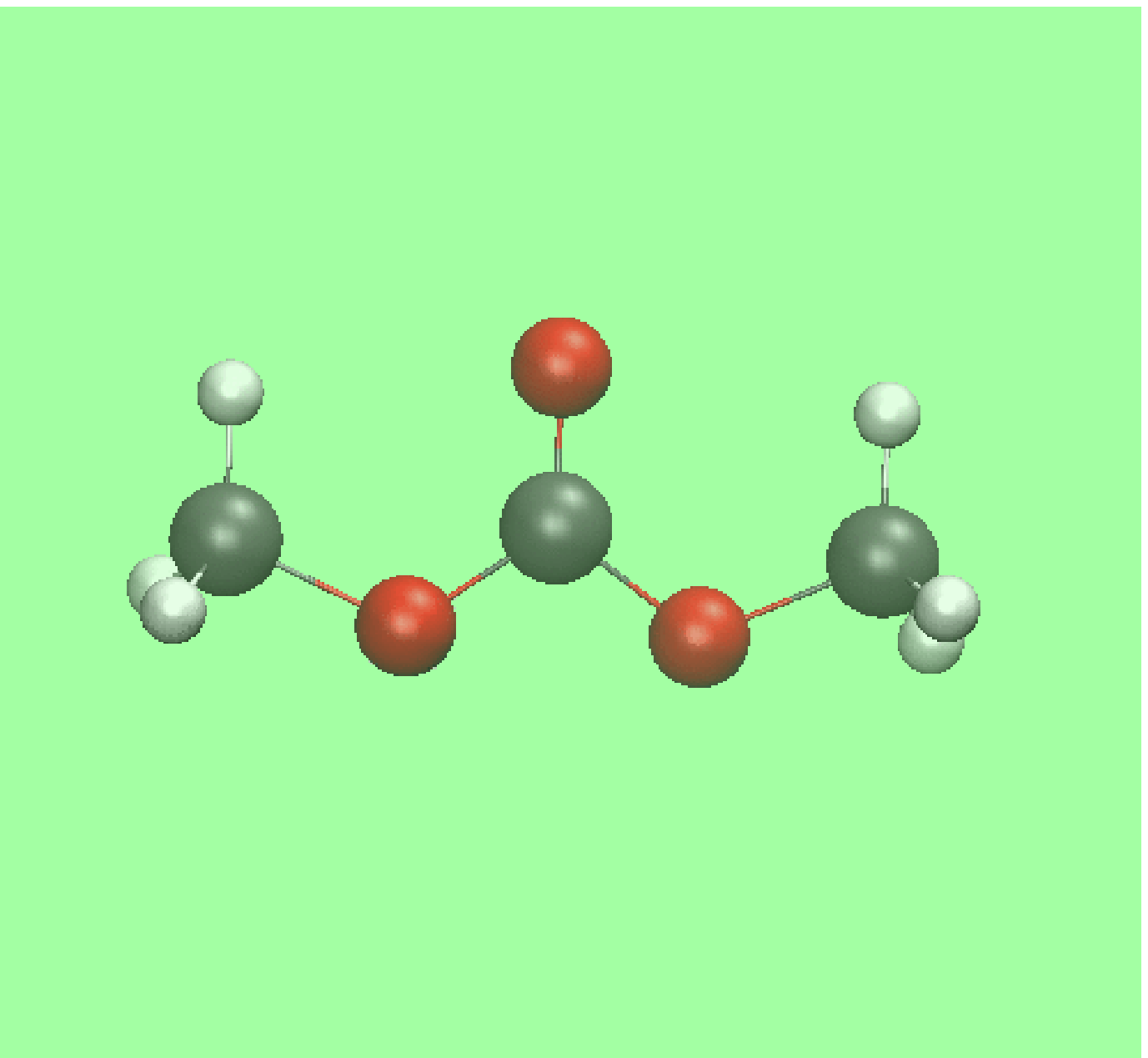} }}
\caption[]
{\label{fig1} \noindent
(a) A simple schematic (not an actual calculation) of a lithium ion battery
depicting a graphitic carbon anode, a LiMn$_2$O$_4$ cathode, and ethylene
carbonate (EC)-based electrolyte.  During charging, Li$^+$ moves from the
cathode to the anode, accompanied by electron flow.  A passivating SEI film,
as labeled, starts to grow upon charging at the anode-electrolyte interface.
Electrolyte oxidation products also emerge on the cathode surface.
(b) EC with atomic labels.  (c) Dimethyl carbonate (DMC).  Grey,
red, white, blue, and purple spheres represent C, O, H, Li, and Mn atoms
respectively.  Liquid state EC are depicted as stick figures.
}
\end{figure}

Lithium ion batteries (LIB) are currently the devices being
implemented or considered for large scale static and transportation
energy storage.  They carry high energy density and have the potential of
dramatically reducing green house gas emission because they operate within
a high voltage window.\cite{book2}  Today's commercial LIBs (Fig.~\ref{fig1}a)
consist of graphitic carbon anodes, transition metal oxide cathodes, and
organic solvent-based electrolyte.  Other crucial LIB components include
passivating ``solid electrolyte interphase'' (SEI) films formed from excess
electron-induced electrolyte decomposition products on anode
surfaces.\cite{book,review,novak_review}  SEI films are heterogeneous
in structure and consist of Li$_2$CO$_3$, ethylene dicarbonate (EDC),
oligomeric/polymeric compounds, salt decomposition fragments, and other
products.\cite{book,review,novak_review}  They prevent continuous electron
injection into the electrolyte, averting further loss of Li$^+$ and
electrolyte molecules.  Li$^+$ transport
through SEI films remains adequately fast.  Oxidation products are also often
found on cathode surfaces.  New concepts of electrodes being pursued, such as
Si-based anodes and ``air'' cathodes in metal-air batteries, share many
solid-liquid interface features shown in Fig.~\ref{fig1}a.  LIBs are pragmatic
devices.  They combine our best electrochemical, solid, and liquid state
expertise to deliver high volumetric and gravimetric energy/power
densities.  While all-solid batteries have received much attention for niche
applications and all-liquid flow-cell batteries have significant potential
for static storage, batteries featuring both liquid and solid components
will undoubtedly dominate for the foreseeable future.

It has been widely acknowledged that interfaces are critical for good 
performance and long lifetime in batteries.\cite{bes}  To some extent,
interfaces dictate the choice of electrode materials and electrolytes.
Graphitic anode forms stable SEI with ethylene carbonate (EC), not propylene
carbonate.\cite{review}  Another celebrated example of the 
interconnected nature of LIB degradation concerns spinel lithium manganese
oxide cathodes.  Mn(II) ions dissolve from the spinel\cite{thackeray_review}  
and diffuse to the anode.  They are incorporated into the SEI there and
degrade its passivation properties via mechanisms not yet fully understood,
leading to enhanced reductive decomposition of the electrolyte.\cite{blyr1998}
Thus solid and liquid degradation modes can be strongly coupled.  The choice
of carbon materials also strongly influences their viability as
anodes.\cite{novak_review}  This is possibly related to interfacial effects
such as instability towards exfoliation of graphite induced by solvent
intercalation\cite{md} and carbon-edge functional groups.  Molecular additives
like vinylene carbonate (VC) have been added to improve electrode
passivation,\cite{review} and these can react on both cathode and anode
surfaces.\cite{dahn} Interfaces are particularly pertinent to nanostructured
electrodes for energy storage applications where the large surface areas call
for enhanced stabilization.\cite{nees1,nees2} If adequate control of
interfaces can be attained, nanostructures have the potential of achieving
much higher electron and lithium ion transport rates, and alleviating
strain-induced electrode cracking and degradation.

To some extent, theory and experiments used to study LIB
liquid-solid interfaces are complementary.  Experimental techniques that
have been applied to SEI studies include Fourier Transform infrared
spectroscopy,\cite{li_vs_lic6} electrochemical impedance
spectroscopy,\cite{li_vs_lic6} X-ray photoelectron spectroscopy
(XPS),\cite{edstrom,lqchen,kanno} atomic force microscopy and Raman
spectroscopy,\cite{ogumi} nuclear magnetic resonance,\cite{leifer} and
transmission electron microscopy.\cite{nees1}  This article only samples
a few experimental papers and focuses instead
on recent atomic-lengthscale modeling of electrochemical reactions at LIB
electrode/electrolyte interfaces.\cite{pccp,bal11,ald,e2,mno,oakridge2}
(See Ref.~\onlinecite{book} for a comprehensive review of experimental methods
and results up to 2004.)
Modeling reactions involve predicting electron transfer and chemical changes
and typically relies on Density Functional Theory (DFT) based techniques.
Such calculations probe smaller length and time scales than experiments, and
can reveal insights about thermodynamics and kinetics which are difficult to
measure under the non-equilibrium, kinetically-driven SEI formation
conditions.  Theoretical SEI growth mechanistic studies were pioneered by
Balbuena {\it et al.}'s cluster-based work\cite{bal01} (see Ref.~\onlinecite{e2}
for a theoretical overview).  With the continual growth of computational power,
the admittedly costly DFT-based {\it ab initio} molecular dynamic method
(AIMD, discussed below) will be a mainstay of future liquid-solid interface
studies.  AIMD permits inclusion of electrodes and liquid electrolytes in model
systems and adds multi-electron reactions, reactive surface sites, and new
perspectives to Balbuena {\it et al.}'s work.  Other interfacial phenomena,
like the dynamics of Li$^+$ intercalation between electrolyte and electrodes,
are important but arguably must await elucidation of how electrode surfaces
are modified by electrolyte decomposition products, unless the electrode
operates at sufficiently modest voltages to keep its surfaces
pristine.\cite{smith_lifepo4,borodin06}

Batteries are complex systems.  The applied voltage, state-of-charge
(lithium content), temperature, electrode cracking, sweep rate, presence of
conductive carbon materials, native surface films like lithium carbonates on
cathodes, and interference from salts/contaminants all affect interfacial
behavior.  They are challenging to model even on bare oxide surfaces
(i.e., in vacuum).\cite{meng_review,ceder,meng_lico2,qi1,qi2,chan,chan1,ceder1,ceder2,benedek11,persson,ouyang} 
We have adopted a basic science approach.  Our starting points are defect-free
surfaces that can be probed using AIMD methods at finite temperature.  Far
from being routine DFT/AIMD applications, we will show that modeling clean
battery interfaces already requires going beyond state-of-the-art theoretical
capability and is a good example of application-driven fundamental research.  

Modeling liquid/solid interfaces is intrinsically interdisciplinary.  This
review will highlight computational techniques which are trivial to either the
solid or liquid state community for the sake of bridging the gap between them.
Important modeling studies on bare electrode surfaces have been made on
crystal facet effects, surface reconstructions and terminations,
non-stoichiometric compositions, and defects.\cite{meng_lico2,ceder,meng_review,qi1,qi2,chan,chan1,ceder1,ceder2,benedek11,persson,ouyang} 
More effort in this area is needed to elucidate surface structures which
are starting points for interfacial studies.  But the electrolyte is
undeniably important and has been somewhat neglected.  To take an example,
DFT-computed energy dffererences for Li insertion into bulk cathode materials 
from Li metal have been reported as ``open circuit voltages''
(OCV),\cite{meng_review} which appears a misnomer (``intrinsic potential''
would be more appropriate).  Measuring OCV invariably involves immersing an
electrode into a liquid electrolyte.  The absolute values of such voltages
are modified by net surface charges, interfacial dipoles, and even purely
quantum mechanical interfacial
effects.\cite{dipole1,quantum,pratt92,lynden1}  Even if interfacial voltage
effects are small in magnitude for battery electrodes, they may become crucial
for atomic level study of Li$^+$ insertion and SEI formation.  We will present
other examples where explicit depiction of molecules qualitatively alters our
understanding of battery degradation processes.  Current theoretical studies
on organic solvent-based Li-air batteries\cite{liair} have also focused on
either the solid\cite{siegel,ceder3,xu_oak} or liquid\cite{solvent} state.

Modeling reactions at LIB electrode/electrolyte interfaces is arguably a
fledging, specialized area.  However, it is part of the broader field of
computational and theoretical electrochemistry, intimately connected to aqueous
interfaces,\cite{gaigeot,shen1,richmond} notwithstanding the fact that water
itself must be excluded from LIB.  DFT-based computational studies of
water-material interfaces pertinent to
electrocatalysis,\cite{halley1988,halley1998,grosse1,grosse2,norskov} 
energy conversion
(water splitting),\cite{selloni_rev,selloni2,michaelides,kubicki,buffalo,car,lynden1} 
geochemistry and mineral dissolution,\cite{benedek12,feooh,sprik1,ugliengo}
and fuel cells\cite{spohr,paddison,otani} can inform and be informed by LIB
simulations.  Our work in fact owes part of its motivation\cite{mno} to water
dissociation studies on TiO$_2$ surfaces.\cite{selloni_rev,selloni2,kubicki,michaelides}
In that example, the interactions of a sub-monolayer of water with different
facets of anatase and rutile have been considered in joint DFT and
experimental studies on clean surfaces under UHV
conditions\cite{selloni_rev,selloni2,michaelides} before AIMD studies of
water-TiO$_2$ interfaces are conducted --- including simulations with anions
which may not be stable under unsolvated UHV conditions.  We argue that a
similar recipe may benefit fundamental studies on LIB.

At the same time, it must be stressed that any frozen (i.e., UHV-like geometry
optimization at T=0~K) description of electrolyte molecules is an uncontrolled
approximation of liquids, although it can yield very useful
insights.\cite{dielectric} Theoretically, one usually claim that
$A$$\approx$$B$ if a low order perturbative expansion between $A$ and $B$
yields accurate predictions.  Liquids and crystalline solids are separated
by discontinuous first order phase transitions.  Elementary calculus dictates
that no real perturbative series connect them.  In that sense, liquids and
crystalline solids are not ``similar;'' they are {\it profoundly} different.
As an example, salt solubility drops many orders of magnitude across the
freezing point -- yet salts are vital to electrochemistry!  Solvent
oxidation/reduction involves changes in solvent charge states, and the
stabilization of these reaction products is related to ionic solubility.
Modeling an electrode with a net charge is facilitated by
explicit treatment of liquid electrolytes that compensate the charge.
AIMD with explicit molecular description of liquids is rigorously suited for
modeling salt effects at interfaces, although this remains costly at present.

This article is organized as follows.  Sec.~\ref{anode} and~\ref{metal}
describe electrolyte decomposition reactions on pristine graphite and
lithium metal surfaces at the initial stages of SEI formation, respectively.
We observe two-electron induced ethylene carbonate decomposition via breaking
two different C-O bonds, instead of the one-electron pathway which has
been the focus in the literature.  Sec.~\ref{ald} discusses electrode coated
with an insulator, where the long-range electron tunneling rate is found to
depend on both the molecular species being reduced and its charge state.
Sec.~\ref{cathode} reviews solvent decomposition modeling on spinel manganese
oxide cathode surfaces and suggests that solvent breakdown and Mn
dissolution may be related.  Sec.~\ref{conclusion} summarizes the article.
A supporting information (S.I.) document is included to discuss computational
challenges to be overcome and possible new directions of research, including
the prediction and control of voltages, electron transfer, and DFT accuracy
issues.  In each section, we briefly describe the motivations and main
predictions at a model LIB interface, discuss the significance in some
detail, and bring up pertinent computational issues.

\section{Liquid EC on LiC$_6$ edges: SEI formation}
\label{anode}

LIB charging occurs at almost $-3$~V vs.~standard hydrogen electrode
(almost 0~V vs.~the Li$^+$/L(s) couple) with anodes like graphite and silicon.
Most electrolyte molecules are electrochemically reduced at such voltages.  On
graphitic carbon, if EC (Fig.~\ref{fig1}b) is a main component of the
electrolyte, stable passivating SEI films are formed from electrolyte
decomposition products, and they prevent further electron leakage
to the electrolyte which causes its continuous breakdown.\cite{review}  This
is somewhat analogous to passivation of some metal surfaces by oxide or
hydroxide in contact with water.  Given its critical importance and ubiquity,
it is tempting to call EC the ``new water'' of lithium ion batteries.

Experimentally, the battery is assembled without pre-intercalation of
Li into graphite.  During first charge, as the voltage is lowered
(effectively putting a negative charge on the anode compensated by
cations at the interface), SEI starts to form.  At even lower voltages, Li$^+$
begins to insert.  This dynamical process is difficult to simulate partly
because the excess charge/voltage relation is difficult to control.  Long
trajectories are needed to equilibrate and converge even the open circuit
voltage drop between the interior of the electrode and the electrolyte
region outside the double layer;\cite{smith_lifepo4} such a calculation may
be slightly beyond current AIMD time scale,\cite{oakridge2} especially
for the purpose of equilibrating salt diffusion.  Modeling
an {\it applied} voltage in a completely condensed-phase setting is
even more challenging.  See the S.I.~for a more detailed discussion.

Instead, we have chosen to start with fully lithiated LiC$_6$ with a
charge-neutral simulation cell (apart from a positive charge arising
from one Li$^+$ ion in the electrolyte region).  While this LiC$_6$
stoichiometry does not truly reflect the experimental conditions during
initial charging, it allows us to set the Li chemical potential of LiC$_6$ in
the solid state anode to the solid state value by approximately
matching the energy of the last Li atom added to the model electrode, in the
absence of electrolytes, with the chemical potential of Li in bulk LiC$_6$.
Unlike on inert electrodes, the voltage in the anode interior can be tuned by
varying Li content without inducing a net charge, although the anode surface
may still retain a new charge; the potential of zero charge has not been
determined experimentally or theoretically for most intercalation charge-neutral
materials.  Thus, with this approach, there remains some ambiguity in the
voltage due to liquid-solid interfacial effects (see the S.I.).  In the next
section, we show that the electrochemical reactions and products discussed in
this section are general; they are also observed at Li metal interfaces, where
(unlike graphite) one does not have to worry whether the SEI forms before
Li intercalation.

\begin{figure}
\centerline{\hbox{ (a) \epsfxsize=1.50in \epsfbox{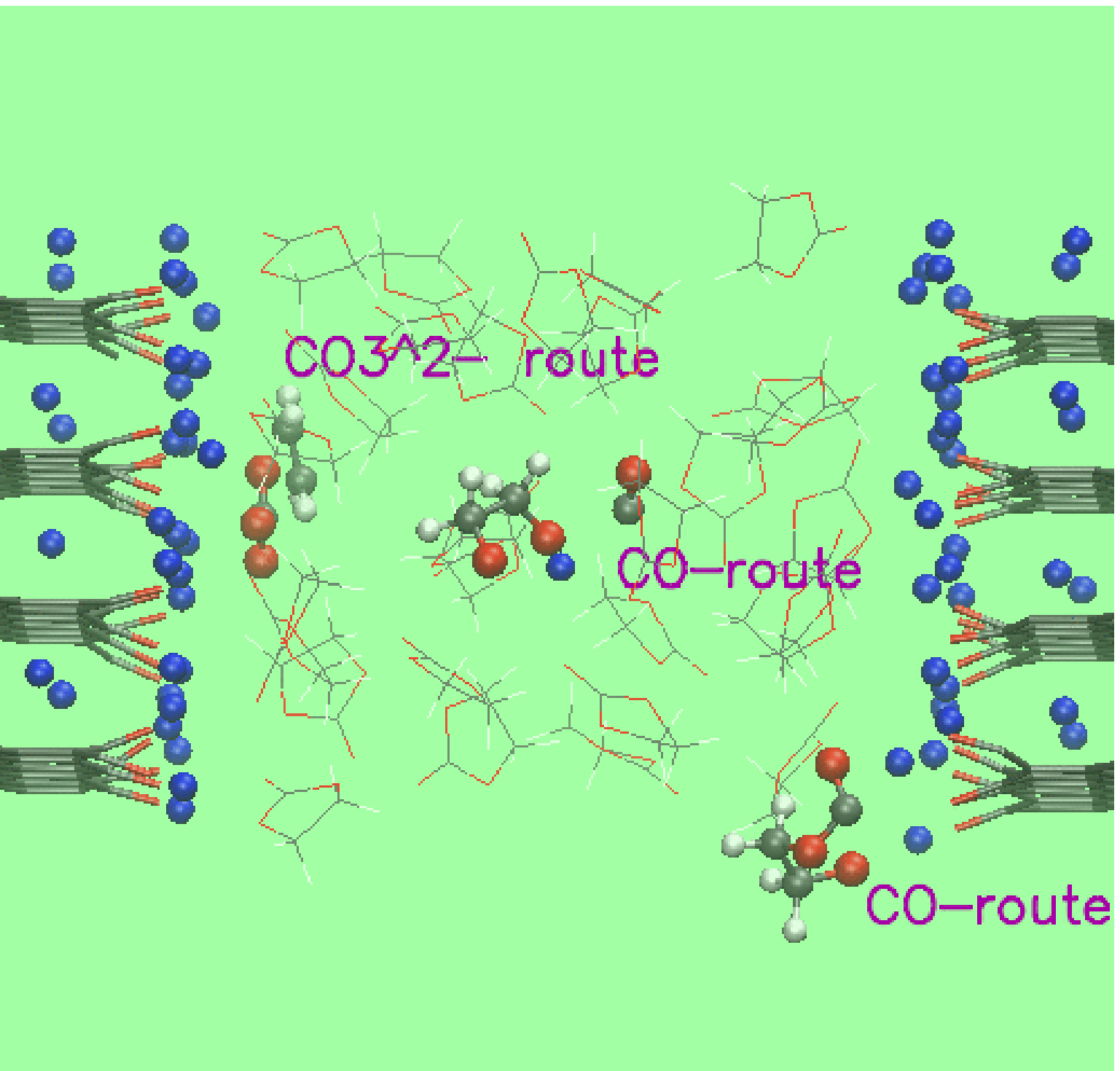}
	           (b) \epsfxsize=1.50in \epsfbox{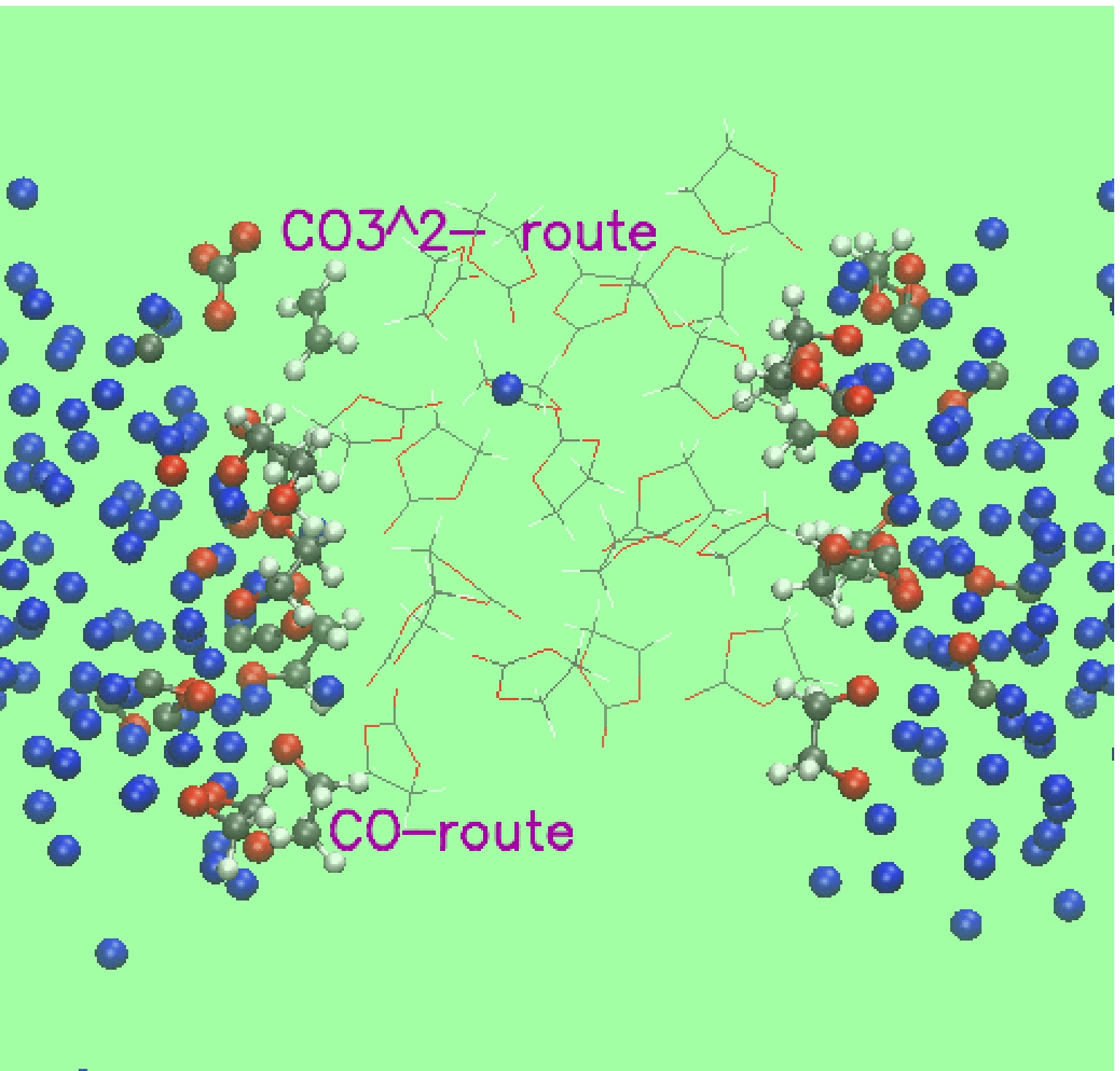} }}
\centerline{\hbox{ (c) \epsfxsize=1.50in \epsfbox{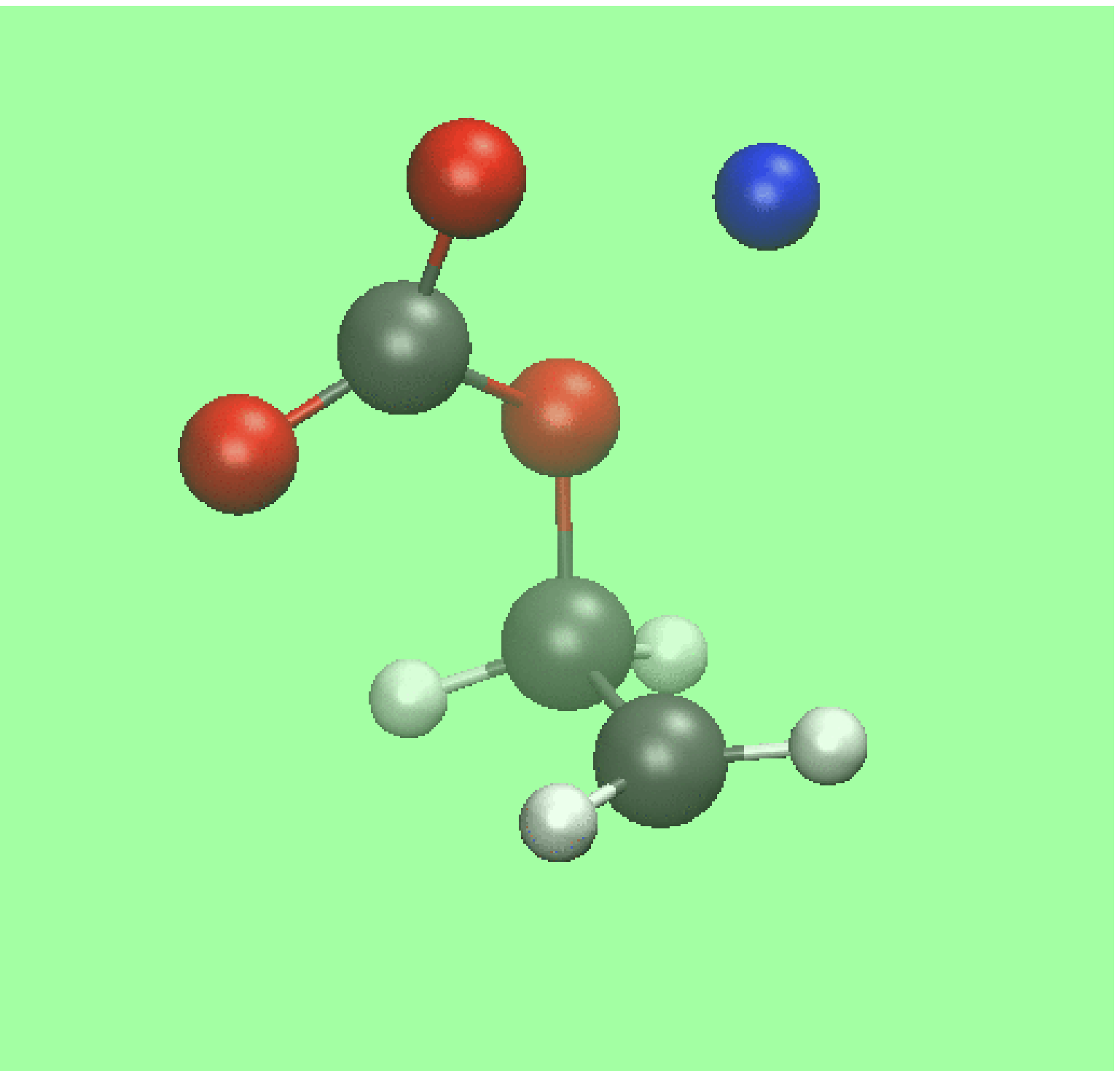}
	           (d) \epsfxsize=1.50in \epsfbox{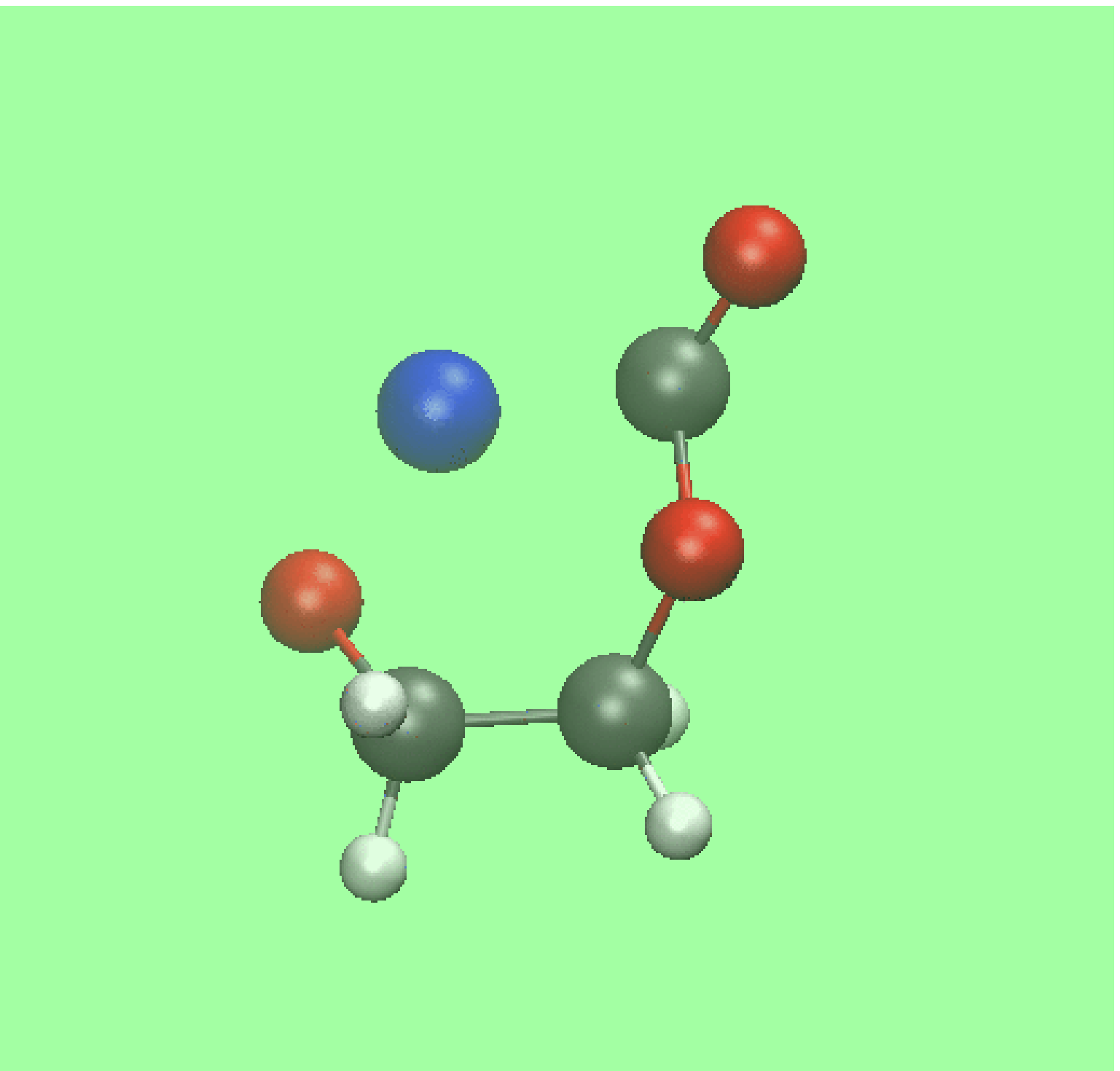} }}
\caption[]
{\label{fig2} \noindent
(a) EC liquid with a solvated Li$^+$ ion confined between pristine C=O
graphite edges after a 7~ps AIMD trajectory.  Intact EC molecules are
shown as wireframes and graphite sheets are depicted as stick figures.  
(b) EC liquid confined between Li metal surfaces after a 10~ps trajectory.
Eleven instances of CO-route decomposition and one of CO$_3^{2-}$ route are
observed.  (c) EC$^-$:Li$^+$ with a broken C$_{\rm E}$-O bond.  Adding a
second $e^-$ to this species yields C$_2$H$_4$ and CO$_3^{2-}$, while
recombination of two ring-opened EC$^-$ radicals can form butylene dicarbonate.
(d) EC$^{2-}$:Li$^+$ with a broken C$_{\rm C}$-O bond; this is the weaker
(lower barrier) bond for EC with two excess $e^-$ and its cleavage leads to
the CO-route.  (c) and (d) are optimized using a dielectric continuum
approximation.\cite{e2}
}
\end{figure}

Figure~\ref{fig2}a depicts an AIMD snapshot at the initial stages of EC
decomposition on graphite pre-intercalated with Li.\cite{pccp} 32~EC molecules
are sandwiched between the anode surfaces.  They are pre-equilibrated using
non-reactive classical force fields.  7~ps into the AIMD trajectory, two
electrons from the electrode have been transferred to each of three EC
molecules coordinated to Li$^+$ ions in bulk solution or at the graphite
edge.  These ``EC$^{2-}$'' species decompose by breaking C$_{\rm E}$-O
and C$_{\rm C}$-O bonds (Fig.~\ref{fig2}a) into two sets of products,
respectively:
\begin{eqnarray}
{\rm EC} + 2 e^- &\rightarrow& {\rm CO}_3^{2-} + {\rm C}_2{\rm H}_4 ;
        \label{old2e} \\
{\rm EC} + 2 e^-&\rightarrow& {\rm OC}_2{\rm H}_2{\rm O}^{2-}+ {\rm CO} .
                        \label{eg-2h}
\end{eqnarray}
Consistent with these reactions, significant amount of C$_2$H$_4$ and
CO gases have been detected in gas chromotography measurement during battery
charging,\cite{yoshida,onuki,ota2,shin} although the precise gas speciation
varies with experimental groups.\cite{novak1998,marom}  Carbon labeling
techniques have shown that EC is a significant source of CO.\cite{onuki}
CO$_3^{2-}$ is known to be a SEI component.  OC$_2$H$_4$O$^{2-}$ is reactive
and can form other products, including oligomers which can be further
reduced.\cite{e2,lee2000}

These AIMD predictions are significant for the following reasons.  In the
literature, the slower one-electron, C$_{\rm E}$-O bond breaking decomposition
route has been much quoted.  It has been invoked to explain most observed
SEI products from ethylene dicarbonate (EDC, reportedly a major SEI
component\cite{edc}) to oligomers.\cite{gewirth,tarascon} Modeling efforts,
including DFT cluster calculations\cite{bal01} and reactive force-field
construction,\cite{shenoy} have also focused on this 1-e$^-$ C$_{\rm E}$-O
cleavage route.  Even if 2-$e^-$ processes are discussed, as they should be
at the initial stages of SEI growth when $e^-$ transfer is fast, most
published works refer to mechanisms that involving breaking C$_{\rm E}$-O
bonds via Eq.~\ref{old2e}.  CO gas release (Eq.~\ref{eg-2h}), reported in
several measurements\cite{yoshida,onuki,ota2,shin} and consistent
with breaking C$_{\rm C}$-O bonds (Fig.~\ref{fig2}d), have arguably been
somewhat ignored.  Because our model system contains a large electron
reservoir and many EC molecules, it supports multiple EC reduction reactions
and decomposition products anchored at the electrodes.  Our unbiased AIMD
simulations, which do not dictate {\it a prior} pathways, predict both
2-$e^-$ induced reactions (Eq.~\ref{old2e}~and~\ref{eg-2h}) at this
out-of-equilibrium, highly driven initial stage of SEI formation.
Thus C$_{\rm C}$-O bond-breaking (Eq.~\ref{eg-2h}, Fig.~\ref{fig2}d) is 
at least as fast as C$_{\rm E}$-O (Fig.~\ref{fig2}c) cleavage.  This has
been confirmed in a quantum chemistry cluster-based publication\cite{e2}
which also suggests that 1-$e^-$ processes only dominate when the SEI thickens
and $e^-$ transfer significantly slows down.  A central insight
of AIMD and AIMD-inspired studies is that the C$_{\rm C}$-O bond becomes
very weak after EC has absorbed one or two electrons.  Therefore Eq.~\ref{eg-2h}
needs to be considered when interpreting experiments and in force field
constructions.  We believe the 2-$e^-$, CO-gas route is a more logical
mechanism to yield EDC, reportedly a main SEI component, than 1-$e^-$
mechanisms.\cite{e2}

We reiterate that the voltage drop between the model electrode and electrolyte
regions is not precisely controlled in these simulations, unlike in experiments.
However, the intrinsic reduction voltage of intact EC$^-$ has been predicted
to be less negative than that of charge neutral EC within the accuracy allowed
by the dielectric continuum approximation used therein.\cite{e2}  This suggests
that the two-electron route and can occur at any applied voltage where EC$^-$
is formed.  The precise applied voltage may still alter the electron transfer
rate via Marcus theory and indirectly change the SEI product
distribution.\cite{e2}

{\it Computational aspects:}
AIMD simulations involve solving, in real time, Newton's second law of
motion ${\bf F}=m{\bf a}$ using forces generated by DFT.  By the ergodic
hypothesis, a sufficiently long trajectory visits all pertinent liquid state
configurations and allows sufficient sampling of equilibrium properties.
A 10~ps AIMD trajectory is at least 10-100 times more costly than zero
temperature optimization calculations for models of the same size.  But
when a finite temperature, explicitly liquid-state, all-molecule description
of electrolyte is used, MD is the rigorous method.  AIMD allows the use
of theoretical models with the highest fidelity to experimental conditions
to date.

These AIMD simulations are based on the approximate DFT/PBE functional\cite{pbe}
which tends to underestimate reaction barriers.  Despite this, they
suggest that much faster 2-$e^-$ reactions will be observed than previous
studies of 1-$e^-$ processes have suggested.\cite{bal01}  Our simulations
of EC are conducted at T=450~K.  One reason is that EC is a solid at room
temperature.  In batteries, the presence of co-solvents like DMC
(Fig.~\ref{fig1}c) reduces the melting point and viscosity.  DMC
{\it cis}-{\it trans} isomerization is currently beyond AIMD time scale, 
but advanced sampling technique may circumvent this problem in the future.
In general, AIMD may not reproduce the correct temperature scale
for liquid state structures and dynamics.\cite{bal11,oakridge1}  In the
celebrated case of liquid water, AIMD/PBE simulations at an elevated T=400~K
is needed to yield T=300~K experimental water structure.\cite{dft_water}

\section{EC on Li metal surfaces: AIMD and UHV modeling}
\label{metal}

EC decomposition products on Li metal surfaces and on LiC$_6$ are qualitatively
similar, although product compositions and salt effects differ in quantitative
ways.\cite{li_vs_lic6}  We have also conducted AIMD simulations of the liquid
EC/Li metal (100) interface (Fig.~\ref{fig2}b).\cite{bal11,ald}  Here EC
reduction reactions are even more violent.  Within 10~ps, all 12 EC~molecules
touching the Li metal have reacted.  11 of the 12 release CO molecules react
via Eq.~\ref{eg-2h} and one EC has decomposed into CO$_3^{2-}$ and C$_2$H$_4$
(Eq.~\ref{old2e}).  Despite the use of a thermostat in the simulation, the
heat generated from the reactions has melted the small Li model electrode.

One signficance of this study is that the open circuit voltage (OCV) should be
unambiguously that of Li$^+$/Li(s).  In contrast, in Fig.~\ref{fig2}a, the
voltage is estimated using solid state approximations, neglecting interfacial
effects.\cite{pccp}  Computing the OCV remains a major challenge of
computational electrochemistry (see the
S.I.),\cite{halley1988,halley1998,grosse1,grosse2} although perturbative
approaches (e.g., adding voltage-induced changes in Fermi levels for 
metallic electrodes) have been successfully applied at T=0~K.\cite{norskov}
Despite this ambiguity associated with the graphite anode, the same SEI products
are predicted (Fig.~\ref{fig2}a and~\ref{fig2}b), showing the robustness
of previous AIMD predictions.\cite{pccp,ald}

Second, the smaller system size of Li electrodes permits the use of a more
accurate but far more computationally costly hybrid DFT functional, namely
HSE06,\cite{pbe0} in AIMD simulations.  While only sub-picosecond trajectories
are possible with AIMD/HSE06, EC is predicted to break the C$_{\rm C}$-O bond
as before,\cite{bal11} giving us further confidence in AIMD/PBE simulations.  
We have also found that increased Brillouin sampling in this smaller system
does not yield different EC reduction reactions.

Finally, the crystalline Li metal (100) surface enables straight-forward
UHV condition DFT/PBE studies of single molecule EC decomposition.  Using
T=0~K nudged elastic band (NEB) calculations, the CO$_3^{2-}$ route is
found to be more exothermic than the CO route on this surface.\cite{ald}
Both mechanisms are shown to be almost barrierless, consistent with
observation of both reactions in picosecond AIMD/PBE trajectories
(Fig.~\ref{fig2}b).  The preference for the CO-route 
is attributed to kinematic factors.  NEB calculations show that electrochemical
EC decomposition mechanisms are predicted to be viable for single EC molecules
at low temperature on Li(100).  This suggests that future UHV imaging of
sub-monolayer EC at low temperature will be valuable.  (Imaging molecules
at liquid/solid interfaces is far more difficult.)  Note that XPS measurements
have been performed on 6-10~nm thick DMC\cite{xps2} and 10-20~nm thick
propylene carbonate (PC)\cite{xps3} films on Li metal surfaces under UHV
conditions.   Carbonates and alkoxides are observed in DMC films, suggesting
both C$_{\rm E}$-O and C$_{\rm C}$-O bond breaking, consistent with our findings
for EC.  PC differs from EC only by a methyl group.  Alkyl carbonate
products are reported in Ref.~\onlinecite{xps3}.  In our simulations, neither
CO gas nor OC$_2$H$_4$O$^{2-}$ is a final product.  CO is absorbed
into the Li metal, while the end groups of OC$_2$H$_4$O$^{2-}$ are
reactive.\cite{lee2000,e2} Future interpretations of XPS measurements of
EC decomposition may benefit from consideration of C$_{\rm C}$-O bond cleavage.

{\it Computational aspects:}
The NEB method, widely used in solid state physics/materials science under
UHV conditions, is similar in spirit to transition state calculations 
performed using the Gaussian suite of programs.  The transition state obtained
should exhibit only one imaginary frequency along the reaction direction.
Unlike solids at T=0~K, liquid state atomic configurations are {\it not} at
local minima, and the instantaneous vibrational frequency spectrum computed at
any MD snapshot contains many imaginary modes.\cite{stratt}  Therefore NEB
cannot be used in the presence of an explicit liquid component (see
Sec.~\ref{cathode}).  Counter ions like PF$_6^-$ are more difficult to stabilize
under UHV conditions, and are better left to AIMD interfacial simulations
which explicitly provide dielectric solvation.\cite{oakridge2}

\section{Atomic Layer Deposition and Electron Transfer}
\label{ald}

\begin{figure}
\centerline{\hbox{ (a) \epsfxsize=1.50in \epsfbox{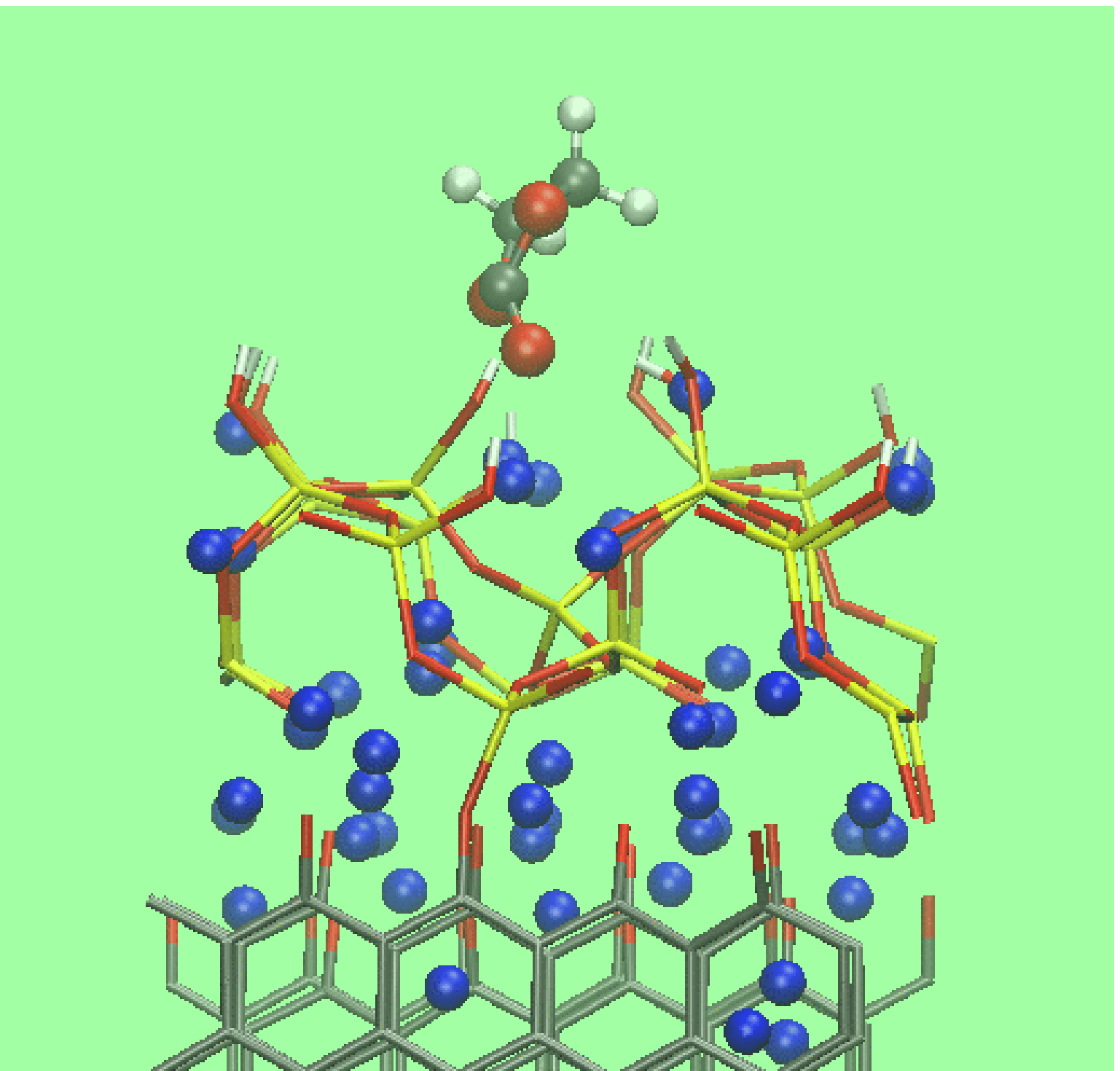}
                   (b) \epsfxsize=1.50in \epsfbox{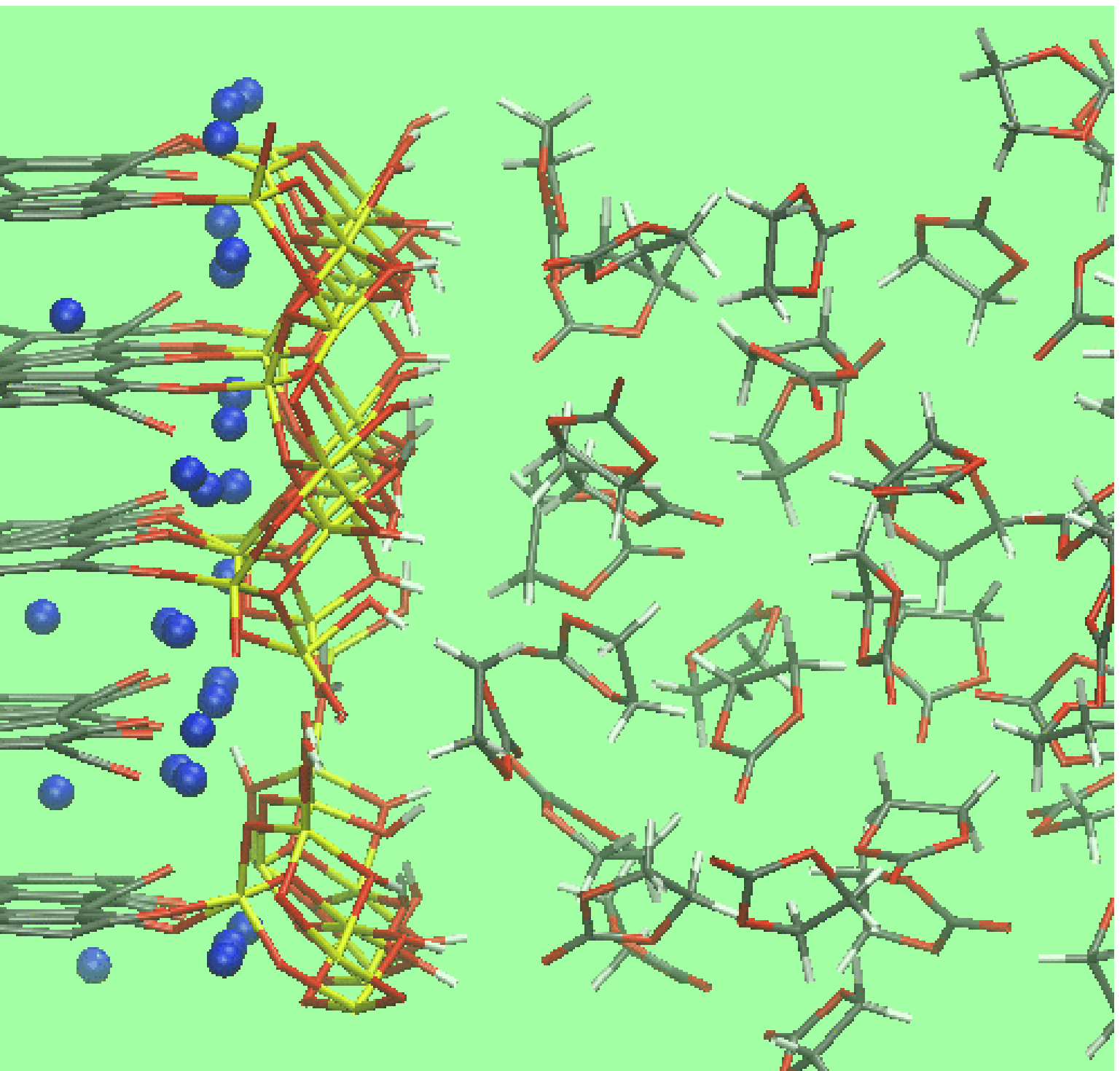} }}
\caption[]
{\label{fig3} \noindent
(a) A single EC$^-$ adsorbed on LiAlO$_2$-coated LiC$_6$ under
ultra-high vacuum conditions.  Electron transfer estimates are made
based on this model.\cite{ald}  (b) EC liquid in contact with
LiAlO$_2$-coated LiC$_6$ strip.  Al is depicted as yellow.
}
\end{figure}

We next consider a LiC$_6$ electrode strip coated with a 7 or 10~\AA\, thick,
hydroxyl-terminated $\beta$-LiAlO$_2$ oxide layer (Fig.~\ref{fig3}).\cite{ald}
These models mimic electrodes coated by the atomic layer deposition (ALD)
technique.\cite{dillon0}  ALD has been used to generate conformal,
sub-nanometer thick oxide layers on electrodes with sub-Angstrom
precision.  This ``artificial SEI'' approach has shown technological
promise in limiting electrolyte decomposition and improving anode and cathode
cyclability.  But ALD oxide layers are also ideal surrogates for insulator
films, including SEI naturally formed from electrolyte decomposition,
in studies of $e^-$ tunneling.

We have estimated the ``non-adiabatic'' electron tunneling rate from the
LiC$_6$ electrode to the adsorbed EC molecule at T=0~K using\cite{marcus,newton}
\begin{equation}
k_{\rm et} = \frac{\sqrt{\pi}|V_o|^2 }{ \hbar
\sqrt { \lambda k_{\rm B}T} }
 \exp \bigg[ -\frac{(\Delta G_o + \lambda)^2}{4 \lambda k_{\rm B}T }\bigg] \, ,
		\label{marcus}
\end{equation}
where $V_o$ is the $e^-$ tunneling matrix element, $\lambda$ is the
reorganization energy, $\Delta G^o$ is $-q_e\Phi$ added to the applied
voltage, and $\Phi$ is the reduction potential.  We set $\Delta G$=0,
which mimics the very initial stage of SEI formation as the applied voltage
is lowed, and estimate that $\lambda$$\approx$2~eV using constrained density
functional theory (cDFT, see below).\cite{voorhis1,voorhis2} $\lambda$ is
found to play a key role in limiting the $e^-$ tunneling rate to about 1/s.  
A rigorous comparison of apparent electron transfer rates associated with
rapid EC decomposition on uncoated anode surfaces (Fig.~\ref{fig2}a \& b)
should in principle be made at the same applied voltage.  However, we
note that C=O and C-OH terminations of LiC$_6$ model electrodes, which
may yield slightly different open circuit voltages, yield similar 
electron-tunneling-induced EC decomposition rates.\cite{pccp}

Ref.~\onlinecite{ald} appears the first theoretical work to use a
Marcus theory-like formulation (Eq.~\ref{marcus}) to study electron transfer
through insulating films on battery anodes.  This work underscores the
importance of depicting explicit molecules.  It is incorrect to think of
$e^-$ tunneling purely as a quantum mechanical barrier-crossing problem, as
though the electrolyte were a featureless $e^-$ sink.  Instead, EC and other
electrolyte molecules/ions exhibit distinct $\Phi$ and $\lambda$
that may lead to vastly different reduction rates.  Perhaps more significantly,
on a per molecule basis, EC is predicted to take on a second electron
(EC$^-$$\rightarrow$EC$^{2-}$) at a rate much faster than the first
reduction event (EC$\rightarrow$EC$^-$),\cite{e2} although much more
rigorous applications of Marcus theory should be used to improve
$e^-$ tunneling rates in the future.  From this insight and using
cluster-based quantum chemistry calculations, we have estimated, speculatively
but apparently for the first time, the crossover between 1-$e^-$ and 2-$e^-$
processes.\cite{e2}  The SEI has often been described as consisting of an
inner, inorganic layer and an outer layer made of organic
carbonates.\cite{book2,book,review}  In
our picture, the inner layer should consist of 2-$e^-$ products, namely
CO$_3^{2-}$ and compounds arising from subsequent reactions of the reactive
OC$_2$H$_4$O$^{2-}$ (Eq.~\ref{old2e} and~\ref{eg-2h}).  The outer layer is
likely butylene dicarbonate arising from barrierless recombination of 1-$e^-$
induced ring-opened EC$^-$ radicals (Fig.~\ref{fig2}c),\cite{e2} but
oligomers\cite{gewirth} and products arising from proton transfer between
EC$^-$ and intact EC molecules may become competitive when $e^-$ transfer
to the electrolyte is slow.

A key difference between SEI growth and aqueous electrochemical reduction
processes is that, in LIB, $e^-$ can be transferred to the solvent molecules
(EC and DMC) right at the electrode surface.  In water,
$e^-$-accepting ionic species are often well-solvated (``outershell'')
complexes located at least Angstroms away from the electrode, and they
experience stronger screening of electric fields by electric double
layers.  However, other electrochemical properties computed using AIMD in
aqueous electrolyes, such as surface potentials\cite{pratt92,surpot} and
reduction potentials,\cite{sprik,ni_paper} are important for organic solvents.
In fact, predicted EC reduction potentials\cite{vollmer} can deviate by tenths
of~eV from the oft-quoted 0.7 to 0.8~V onset observed in experiments on
graphite and even on TiO$_2$ anodes.\cite{tio2_sei}  More accurate predictions
of $\Phi$,\cite{borodin12} especially for EC adsorbed on electrode surfaces,
will improve the predicted $e^-$ transfer rate via the $\Delta G$ term in
Eq.~\ref{marcus}.  Using explicit solvent models to calculate $\Phi$ should
also be considered.

{\it Computational aspects:}  Electron transfer may be ``adiabatic''
or ``non-adiabatic'' (Eq.~\ref{marcus}) depending on $V_o$ and
and $\lambda$.\cite{marcus,newton}  In the former regime, electronic
configuration responds instantaneously to nuclear degrees of freedom.  DFT
calculations assume such a Born-Oppenheimer separation of time scales, and are
appropriate at the initial stages when the electrolyte is in contact
with pristine metallic electrodes.  In the non-adiabatic regime, electron
tunneling is slow, and the system is not necessarily in its ground electronic
state.  DFT is problematic here.  For example, it cannot confine $e^-$ behind
an insulating film if electrons are only metastable on the electrode at a
particular atomic configuration.

Thus ALD-coated electrodes present a challenging prototype problem for
DFT-based studies.  We have found that T=0~K DFT NEB calculations based
on the widely used PBE functional allows the unphysical splitting of an
electron between the electrode and the EC molecule without a large energy
cost.\cite{ald}  This appears to be an example of self-interaction error
(SIE).\cite{wtyang} Consequently, the $e^-$ tunneling barrier appears 
underestimated using this functional.  The problem may be alleviated
using more costly hybrid functionals which are less susceptible to SIE.

We have instead applied a combination of PBE calculations and cDFT
which has been revitalized by van Voorhis and others.\cite{voorhis1,voorhis2}
We have implemented local shape functions $f_i(r)$ centered around all
atoms (``$i$'') of the EC molecule, which allows us to approximately add or
subtract one $e^-$ into or from a EC molecule adsorbed on the insulating
LiAlO$_2$ film at T=0~K.  This gives $\lambda$ predictions via relaxation of
vertical electronic excitations,\cite{ald,voorhis1,voorhis2} and yields an
estimate of the non-adiabatic $V_o$ as a by-product.\cite{voorhis2,blumberger}
Preliminary attempts to compute vertical excitations at explicit liquid-solid
interfaces (Fig.~\ref{fig3}b), related to finite temperature $\lambda$
calculations, have been made but they remain costly.  See Ref.~\onlinecite{ald}
and the S.I. for discussions of the approximations used and avenues of
improvement in this important area.

Long-range electron transfer is clearly pertinent to aqueous electrolyte
interfaces where the electrode is coated with an oxide or hydroxide layer.
So far most computational work on electron transfer at water-material
interfaces have focused on pristine metal
electrodes.\cite{grosse1,willard,boroda,schmickler}  Many energy applications
feature non-precious metal electrodes that react with the solvent to form
insulating films.

\section{EC decomposition on cathode surfaces}
\label{cathode}

\begin{figure}
\centerline{\hbox{  \epsfxsize=2.9in \epsfbox{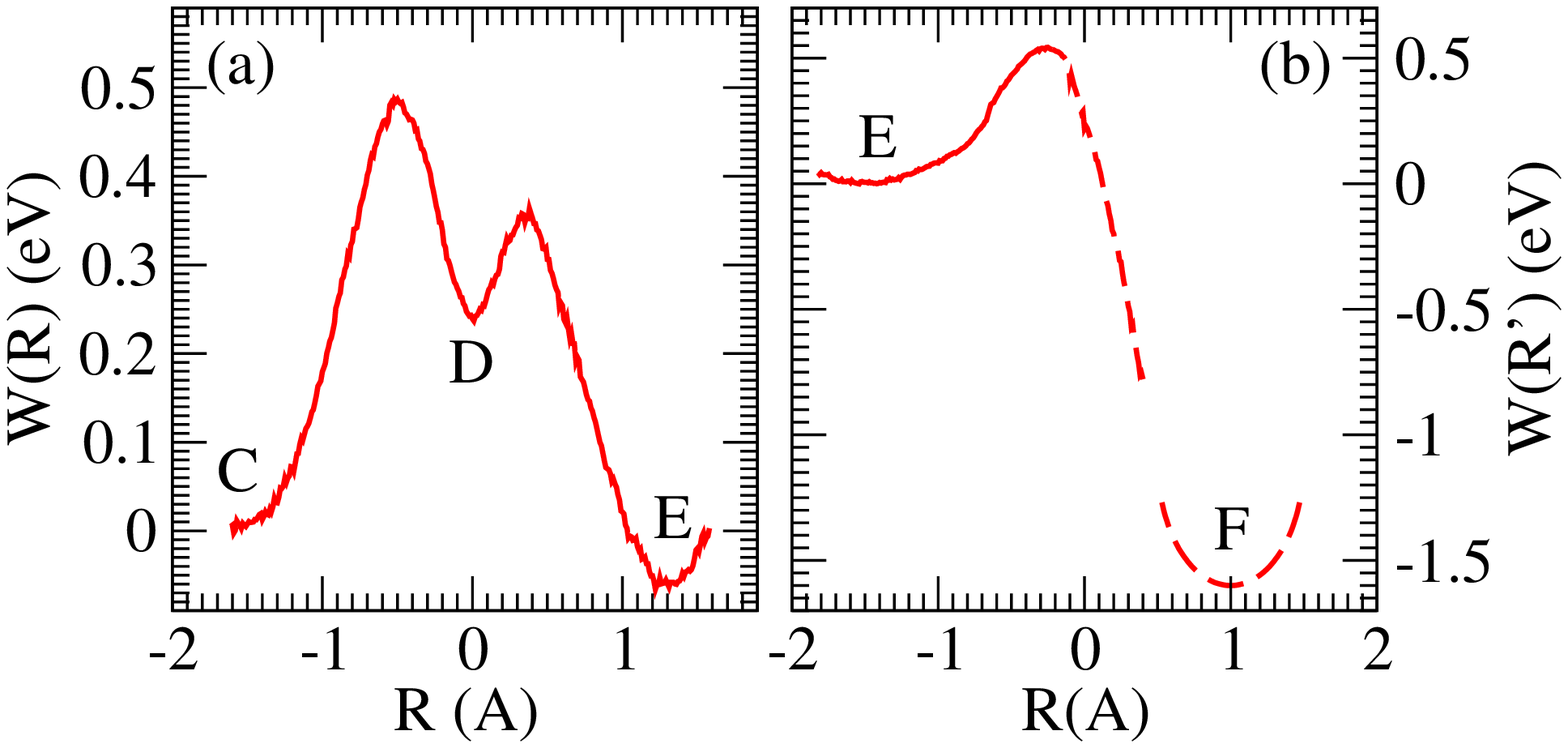} }}
\centerline{\hbox{ (c) \epsfxsize=1.125in \epsfbox{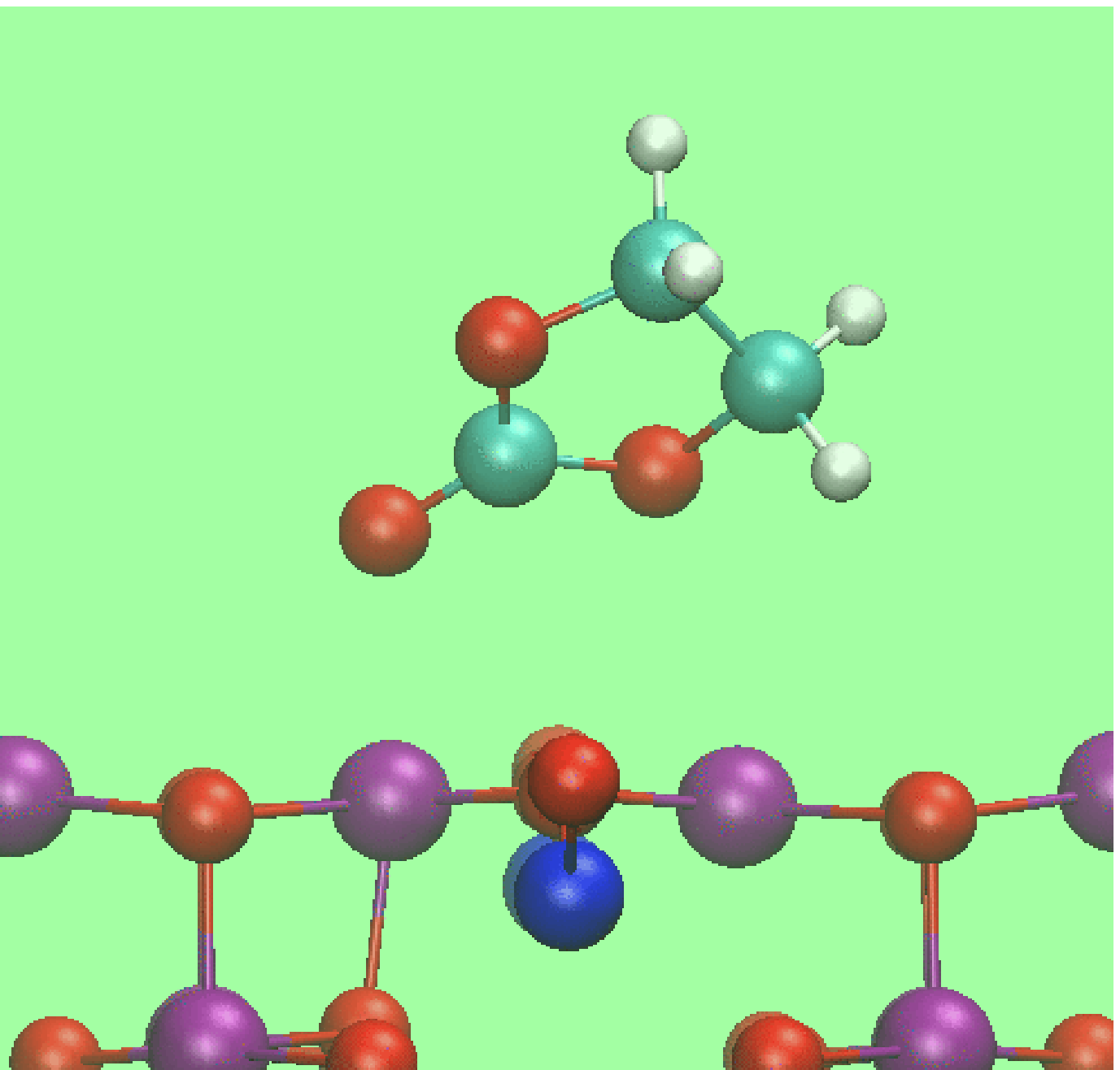}
                   (d) \epsfxsize=1.125in \epsfbox{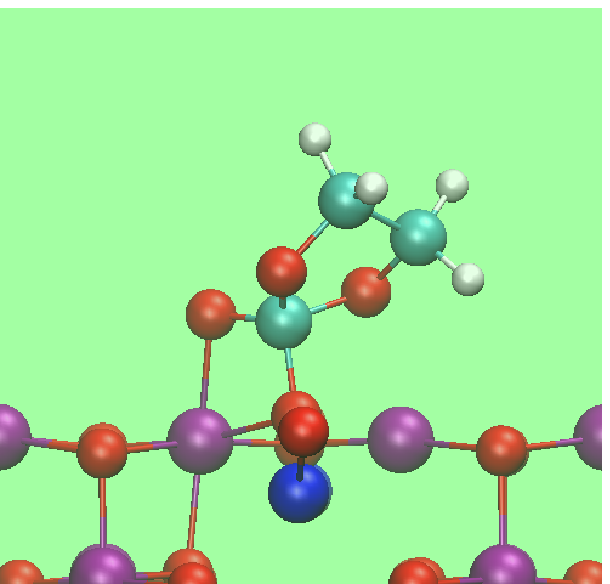} }}
\centerline{\hbox{ (e) \epsfxsize=1.125in \epsfbox{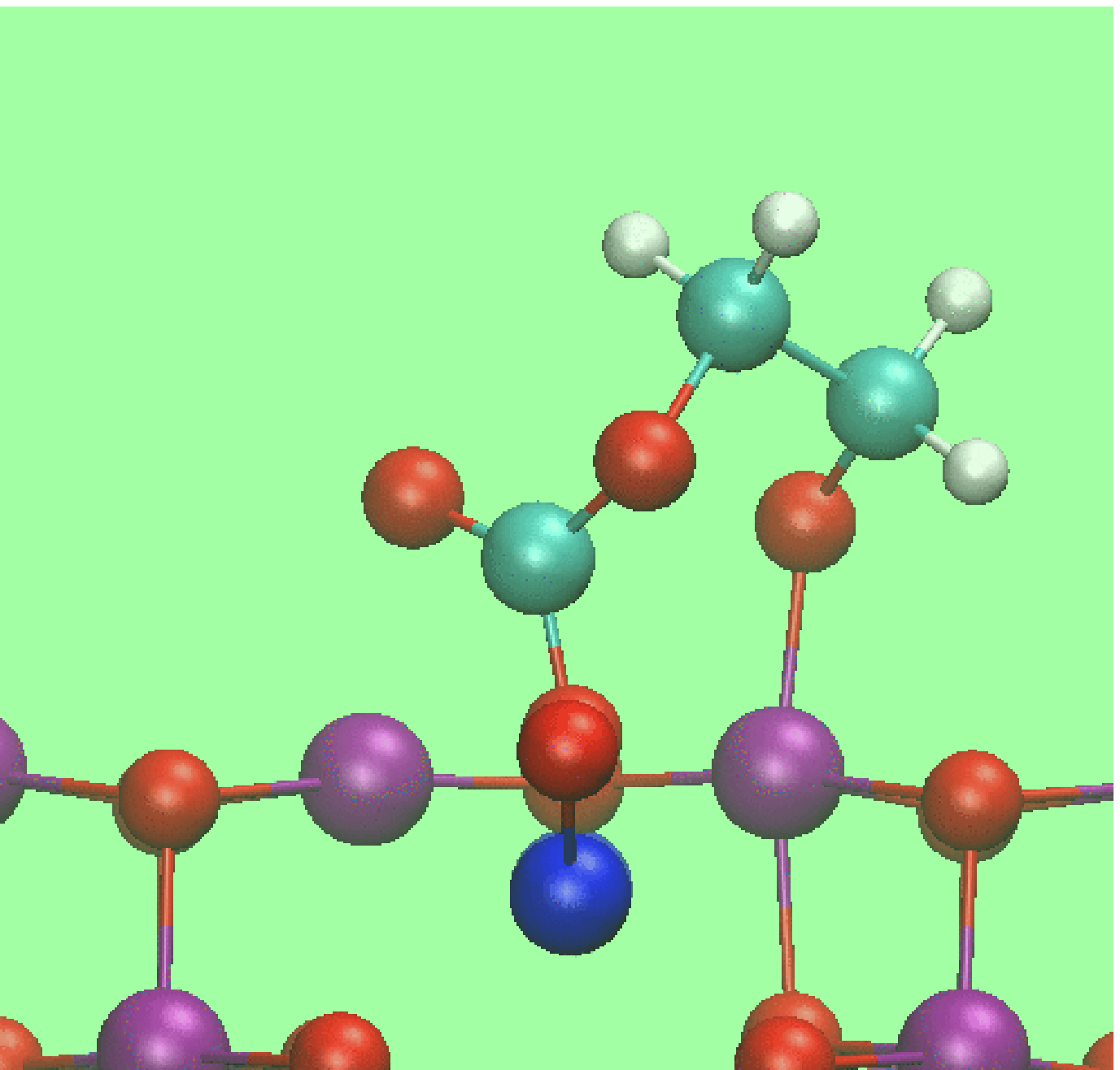}
                   (f) \epsfxsize=1.125in \epsfbox{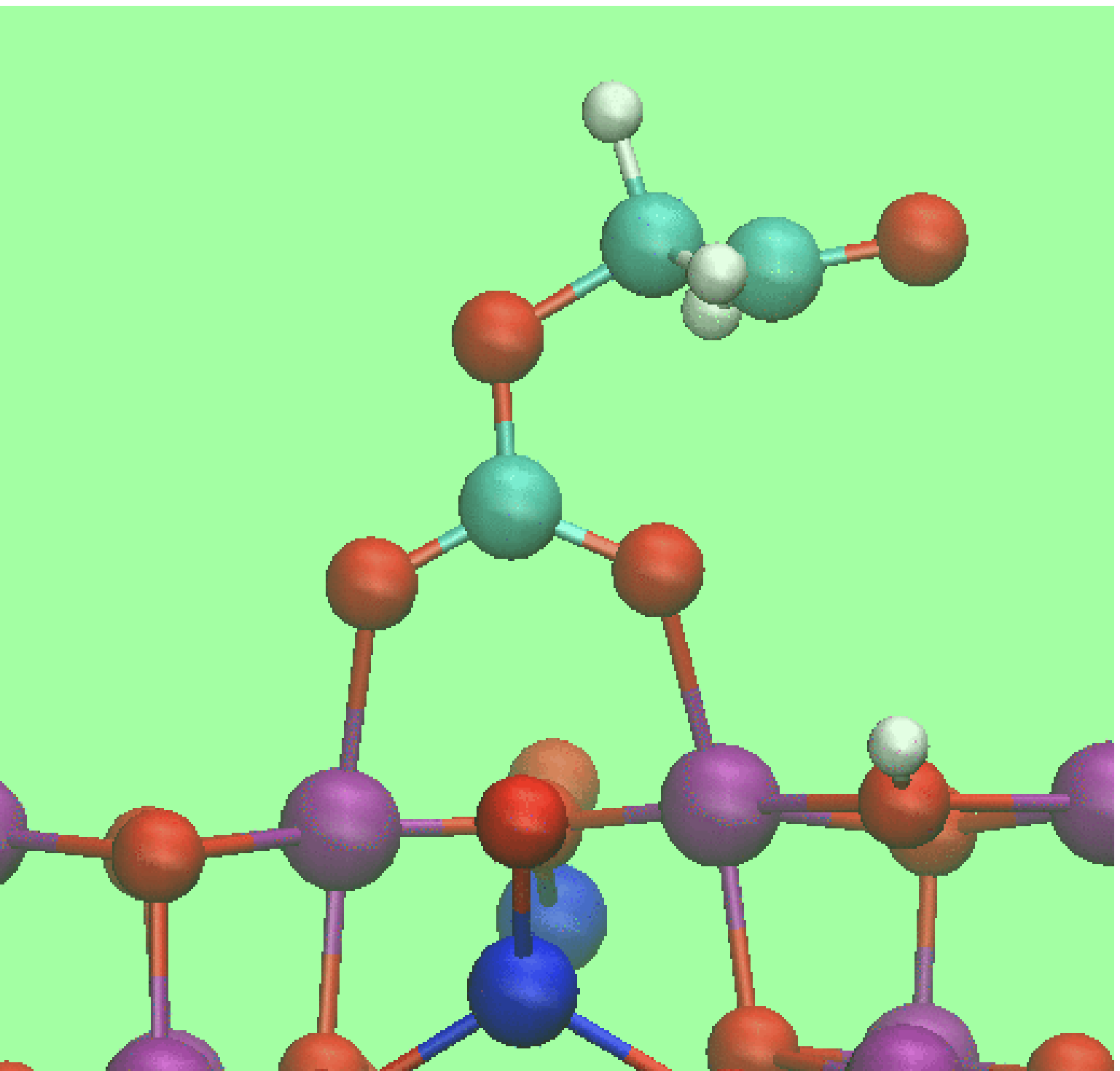} }}
\caption[]
{\label{fig4} \noindent
(a)-(b) Potentials-of-mean-force for two segments of the EC oxidation
reaction.  (c) Intact EC (configuration C) on Li$_{0.6}$Mn$_2$O$_4$
(100) surface.  (d) Intermediate~D.  Note that the EC C$_{\rm C}$ atom
sits atop a surface oxygen ion that is not bonded to a Mn immediately
below.  (e) Intermedate~E, with a broken C$_{\rm C}$-O$_{\rm E}$ bond.
The surface Mn(III) ion coordinated to the O$_{\rm E}$ now becomes
a Mn(IV).  (f) Product~F; a proton and two electrons are transferred
to the surface.
}
\end{figure}

We have also studied EC decomposition on the (100) surfaces of spinel
Li$_x$Mn$_2$O$_4$,\cite{mno} a promising choice of LIB cathode
material.\cite{thackeray_review}  The issues to be addressed are twofold.
First, electrolytes are known to decompose on spinel oxide surfaces despite
the fact that the working voltage range of Li$_x$Mn$_2$O$_4$ is not high.  The
``intrinsic'' oxidation voltage of EC molecules, computed in the absence of
electrodes even under conditions most favorable to oxidation (EC coordinated at
PF$_6^-$ in a low dielectric constant medium\cite{borodin1}), lies outside the
spinel operational window.  This suggests that, for EC to be oxidized, the
surface has to play a catalytic/reactive role.  The adsorbed, oxidized 
organic fragments have not been fully identified.  Second, the spinel
oxide also degrades and Mn ions dissolve from it.\cite{blyr1998} 

To our knowledge, Ref.~\onlinecite{mno} is the first theoretical study of
organic solvent molecule reactions on LIB cathodes.  Unlike on anodes,
unconstrained AIMD trajectories of EC liquid on defect-free
Li$_{0.6}$Mn$_2$O$_4$ (100) surfaces do not lead to spontaneous
electrochemical reactions.  The required timescale is apparently too
long.\cite{meta}  To guide AIMD studies, we first consider a pristine
crystal surface under UHV conditions.  NEB calculations show that chemisorption
of a single
EC with an internal C$_{\rm C}$-O$_{\rm E}$ bond broken and with an EC oxygen
atom strongly bound to a 5-coordinated surface Mn(IV) ion preceeds oxidation.
This chemisorbed geometry leads to a very exothermic transfer of two electrons
and a proton to the oxide surface.  DFT/PBE0-predicted energetics are
similar for the intermediates.\cite{mno}

Then AIMD potential-of-mean-force (PMF, see below) simulations are conducted
on explicit liquid EC/electrode interfaces at finite temperature to investigate
this UHV-motivated pathway.  These simulations yield results similar to UHV
predictions (Fig.~\ref{fig4}).  In the first two steps, no charge transfer
occurs, and the dielectric solvation missing in UHV studies should not play
a strong role.  In the key third step, electrons and a H$^+$ are transferred,
and the liquid environment slightly lowers the reaction barrier compared to 
UHV calculations.  The predicted barrier appears readily surmountable in
battery operation time scale.

One significance of this work is that it suggests solvent oxidation and
electrode degradation may be related.  The oxidized EC molecule fragment
has pulled an oxygen ion out of the surface (Fig.~\ref{fig4}f).  The proton
transfer from EC to the oxide is perhaps even more intriguing.  Acid has
been found to accelerate Mn(II) dissolution.\cite{thackeray_review} Acid
protons have often been assumed to come from trace water,
inevitably present) reacting with the PF$_6^-$ salt.  However, Oh and
coworkers have reported evidence that protons can come from organic solvent,
especially tetrahydrofuran with
LiClO$_4$ salt.\cite{oh}  Our proposed EC oxidation mechanism may take place
cooperatively with the trace water route, because H$^+$ deposited on the oxide
surface may create H$_2$O as Mn dissolves.  This work paves the way for future
studies of oxidation on spinel (111) surfaces, which are more prominent
than (100).  Incidentally, the surface reconstruction of the (111) surface
in vacuum has not been published.\cite{persson}  For liquid-state experts,
this last point emphasizes the difficulty of identifying appropriate starting
surfaces to perform liquid-solid interfacial calculations. Finally,
as-synthesized LiMn$_2$O$_4$ particles are covered with lithium carbonate and
inactive surface oxide phases.  For Mn ions to dissolve, the carbonate film
must first either crack\cite{novak2} or dissolve in the electrolyte.  

Much attention has been paid to Mn charge states, especially on
bare spinel oxide surfaces.\cite{ouyang,benedek11}  This is because Mn(II)
dissolution has been proposed to occur via two Mn(III) disproportionating
into Mn(II) and Mn(IV).\cite{thackeray_review}  However, for an Mn ion to
dissolve directly from the spinel surface, it must first coordinate to one
or more solvent molecules.\cite{benedek12} We have observed that, when an
EC molecule binds to a 5-coordinated surface Mn(III) on the (100) surface
under UHV conditions, the now 6-coordinated Mn(III) transforms into a Mn(IV)
by donating an electron to a subsurface Mn(IV).  This seems to render Mn
charge states on the bare (100) surface irrelevant, and emphasizes that
molecules must be explicitly depicted to understand Mn dissolution.

{\it Computational Aspects:}
As discussed in Sec.~\ref{metal}, NEB cannot be used to find reaction
barriers in the presence of explicit liquid components.  Instead, AIMD
PMF calculations are performed using the umbrella sampling method.\cite{book1}
Reactive coordinates $R$ that continuously link the reactant, transition
states, and product are chosen using UHV results as guide.  The free energy
change along $R$ is proportional to $-k_{\rm B}T$$\ln P(R)$, where $P(R)$ is
the probability that the value $R$ occurs in a trajectory after correcting
for effects of umbrella constraints (Fig.~\ref{fig4}a~\&~b).\cite{mno}
Calculating the reaction free energy ($\Delta G$), which is path-independent,
also benefits from having a continuous $R$ path.  An alternative would be to
calculate the average enthalpies of the initial and final state and subtracting
them.  This alternate approach is plagued by thermal uncertainties which scale
as $(N \tau/t)^{0.5} k_{\rm B}$T, where $N$ is the number of nuclear degrees
of freedom, $t$ is the total trajectory length, and $\tau$ is the correlation
time.  If the simulation is faithful to experiments and features many liquid
molecules (large $N$), the noise can easily be a fraction of an electron
volt.  The PMF method circumvents this problem.

\section{Conclusions}
\label{conclusion}

We have reviewed our modeling work on lithium ion battery electrode/electrolyte
interfacial electrochemical reactions leading to SEI formation on anode
surfaces and the reactions between electrolytes and spinel lithium manganese
oxide cathode surfaces.  AIMD simulations are valuable for predicting unbiased,
fast, kinetically-determined multiple reaction mechanisms at the initial
stages of SEI formation.  Although costly, insights gained using AIMD 
can motivate new research directions which require less computationally
intensive methods.  Electron tunneling between electrode and electrolyte
through an insulating layer will be an important area of research (see also
the S.I.).  The explicit depiction of molecules and electrode surfaces is
shown to be important for both electron transfer and cathode degradation.
The utility of ultra-high vacuum experiments on clean electrode surfaces
covered with sub-monolayers of electrolyte molecules is proposed.  We have
also highlighted the synergy between lithium ion battery modeling and
theoretical studies of water-material interfaces.\cite{gaigeot,shen1,richmond}
Exchange of knowledge between these disciplines will accelerate progress
in modeling battery processes where more fundamental modeling studies are
necessary.

\section*{Acknowledgement}
 
This work is funded by Nanostructures for Electrical Energy Storage (NEES),
an Energy Frontier Research Center funded by the U.S. Department of Energy,
Office of Science, Office of Basic Energy Sciences under Award Number
DESC0001160.  We thank Dr. Ashley Predith and NEES PIs and affiliates for
discussions.  Sandia National Laboratories is a multiprogram laboratory
managed and operated by Sandia Corporation, a wholly owned subsidiary of
Lockheed Martin Corporation, for the U.S.~Deparment of Energy's National
Nuclear Security Administration under contract DE-AC04-94AL85000.

\section*{Supporting information available}
Supporting information on modeling challenges and new directions, including
issues related to prediction and control of voltages, electron transfer,
and DFT accuracy is provided.  This information is available free of charge
via the Internet at {\tt http://pubs.acs.org/}.


\newpage

\section*{Brief biography}

Kevin Leung did his PhD studies in Chemistry at the University of California
at Berkeley, and went on to postdoctoral appointments at the University
of Southern California and in Berkeley before becoming a Sandia staff scientist.
His main interest is chemical reactions in liquids and at liquid-solid
interfaces, {\it ab initio} molecular dynamics, computational electrochemistry,
batteries, and carbon dioxide- and energy-related research.


\begin{references}
\bibitem{marom}
The gas products evolvefd have also been analyzed by battery and car
companies, but much of this information remains proprietory.
\end{references}
\end{document}